\documentclass[aip,pop,reprint]{revtex4-1}
\usepackage{graphicx}
\usepackage{amsmath}
\usepackage{mathrsfs}
\usepackage{array}
\usepackage[breaklinks=true,colorlinks=true,linkcolor=blue,urlcolor=blue,citecolor=blue]{hyperref}
\graphicspath{{./figures/}}

\begin{document}

\title{A Lagrangian model for laser-induced fluorescence and its application to measurements of plasma ion temperature and electrostatic waves} 
\author{F. Chu}
\email[]{feng-chu@uiowa.edu}
\affiliation{Department of Physics and Astronomy, University of Iowa, Iowa City, Iowa 52242, USA}
\author{F. Skiff}
\affiliation{Department of Physics and Astronomy, University of Iowa, Iowa City, Iowa 52242, USA}
\date{\today}

\begin{abstract}
\rightskip.5in
Extensive information can be obtained on wave-particle interactions and wave fields by direct measurement of perturbed ion distribution functions using laser-induced fluorescence (LIF). For practical purposes, LIF is frequently performed on metastable states that are produced from neutral gas particles and ions in other electronic states. If the laser intensity is increased to obtain a better LIF signal, then optical pumping can produce systematic effects depending on the collision rates which control metastable population and lifetime. We numerically simulate the ion velocity distribution measurement and wave-detection process using a Lagrangian model for the LIF signal for the case where metastables are produced directly from neutrals. This case requires more strict precautions and is important for discharges with energetic primary electrons and a high density of neutrals. Some of the results also apply to metastables produced from pre-existing ions. The simulations show that optical pumping broadening affects the ion velocity distribution function (IVDF) $f_0(v)$ and its first-order perturbation $f_1(v,t)$ when laser intensity is increased above a certain level. The results also suggest that ion temperature measurements are only accurate when the metastable ions can live longer than the ion-ion collision mean free time. For the purposes of wave detection, the wave period has to be significantly shorter than the lifetime of metastable ions for a direct interpretation. It is more generally true that metastable ions may be viewed as test-particles. As long as an appropriate model is available, LIF can be extended to a range of environments. 

\end{abstract}

\maketitle

\section{Introduction}
\label{sec:intro}

Plasmas have many electro-mechanical degrees of freedom. In collisionless plasma, for example, one has the continuum of Case-Van Kampen modes \cite{van_kampen_theory_1955, case_plasma_1959}. These degrees of freedom are difficult to observe and analyze through low moments of the ion velocity distribution function (IVDF) such as density and flow velocity. Thus it is important to have an accurate measurement of the IVDF itself and its perturbation to understand plasma response \cite{skiff_electrostatic_2002, skiff_linear_1998}. A reliable phase-space diagnostic is also required in the study of ion heating, velocity-space diffusion and related phenomena in gas discharge, fusion, and other plasmas \cite{anderegg_ion_1986, skiff_direct_1987, curry_measurement_1995}.

Laser-induced fluorescence (LIF) is a nonintrusive, nominally nonperturbative diagnostic technique that has found application in the study of a wide range of fundamental and applied problems. Though it is possible to use laser fields to heat or cool plasma ions \cite{hollmann_measurement_1999}, in the case where the effect of photon momentum on ion orbits is negligible, LIF provides an important window into the dynamics of ion motions in phase-space. LIF is produced from allowed transitions of plasma ions that are optically pumped to excited states. This optical pumping process that is dependent on ion orbits plays a key role in LIF and ultimately is non-linear in laser intensity. An accurate theoretical model of LIF is needed to provide a guideline for avoiding or correcting systematic errors in LIF measurements, such as optical pumping broadening \cite{goeckner_laserinduced_1989, goeckner_saturation_1993, severn_argon_1998} and metastable lifetime effects. 

Laser-induced fluorescence is normally analyzed through an Eulerian approach based on local measurements in the phase-space of position and velocity. In this picture, the rate equations are extended to a system of coupled kinetic equations, with one for each quantum state. This Eulerian model has been used to study optical tagging \cite{skiff_nonlinear_1995}, velocity-space diffusion \cite{curry_measurement_1995}, optical pumping broadening \cite{goeckner_laserinduced_1989}, and many other problems \cite{skiff_mini-conference_2004}. This traditional Eulerian model determines the density distribution and the velocity or energy distribution of plasma ions, but it also has limitations. The solutions of the coupled kinetic equations, being a set of coupled partial differential equations (PDEs), become extremely difficult to compute with the existence of nonuniformity in phase-space. This problem becomes acute if the effects of waves are included. 



To cope with these problems, an interpretation of LIF based on a Lagrangian approach is introduced. In this picture, one must follow each individual ion orbit as it moves through space and time. The approach separates the classical dynamics of the ions from the quantum mechanics of the electronic states. This permits easily extending the validity of the calculation to the nonlinear regime of LIF where the fluorescence signal no longer grows proportionally with laser intensity. The induced transition rate, however, is still typically much smaller than the spontaneous decay rate. The Lagrangian approach provides a large computational advantage, as it reduces the system of coupled PDEs to ordinary differential equations (ODEs). In many cases, the phase-space integrals can be computed analytically; reducing them to functions that can be rapidly evaluated using continued fractions \cite{mccabe_continued_1984}. Some early work has been done based on using the optical pumping produced by a single laser beam on a short time scale in a uniform electric field through this approach \cite{claire_nonlinear_2001}. This paper presents a more general Lagrangian model by introducing a conditional probability function $P(x,v,t;x',v',t')$ valid for long time $\nu(t-t')\gg 1$ where $\nu$ is the ion-ion coulomb collision frequency. Since this model does not impose constraints on the particle orbits, it can be applied to systems with complicated ion dynamics such as the response to electrostatic waves.

Metastable ions can be produced from both neutral gas particles \cite{cherrington_gaseous_1979} and ions in other electronic states \cite{goeckner_laserinduced_1991}. In the Lagrangian approach, the contribution from these two populations can be considered independently and simply summed to provide the total LIF signal. We have two versions of models dealing with these two different situations. As the metastable population from neutrals presents some unique problems (see Sec. \ref{sec:LIF}), in this paper we focus on understanding metastable ions produced directly by single-step ionization of neutrals by electron impact. Nevertheless some of the results also apply to metastables produced from pre-existing ions.


Optical pumping broadening has been theoretically studied before using an Eulerian approach \cite{goeckner_laserinduced_1989}. Here we demonstrate this instrumental effect based on the Lagrangian approach. Results concerning other effects that are difficult to study through the Eulerian approach, such as coulomb collision and metastbale lifetime effects, are also considered here. This paper is organized as follows: Sec. \ref{sec:LIF} presents a typical LIF scheme and rate equations, Sec. \ref{sec:theo} gives a description of the conditional probability function and derivation of the Lagrangian model for LIF, Sec. \ref{sec:exp} gives a description of the experimental setup, Sec. \ref{sec:results} presents the simulation and experimental results, and Sec. \ref{sec:summary} provides a summary.

\section{LIF Scheme and Rate Equations}
\label{sec:LIF}

\begin{figure}
\begin{center}
\includegraphics[width=2.8in]{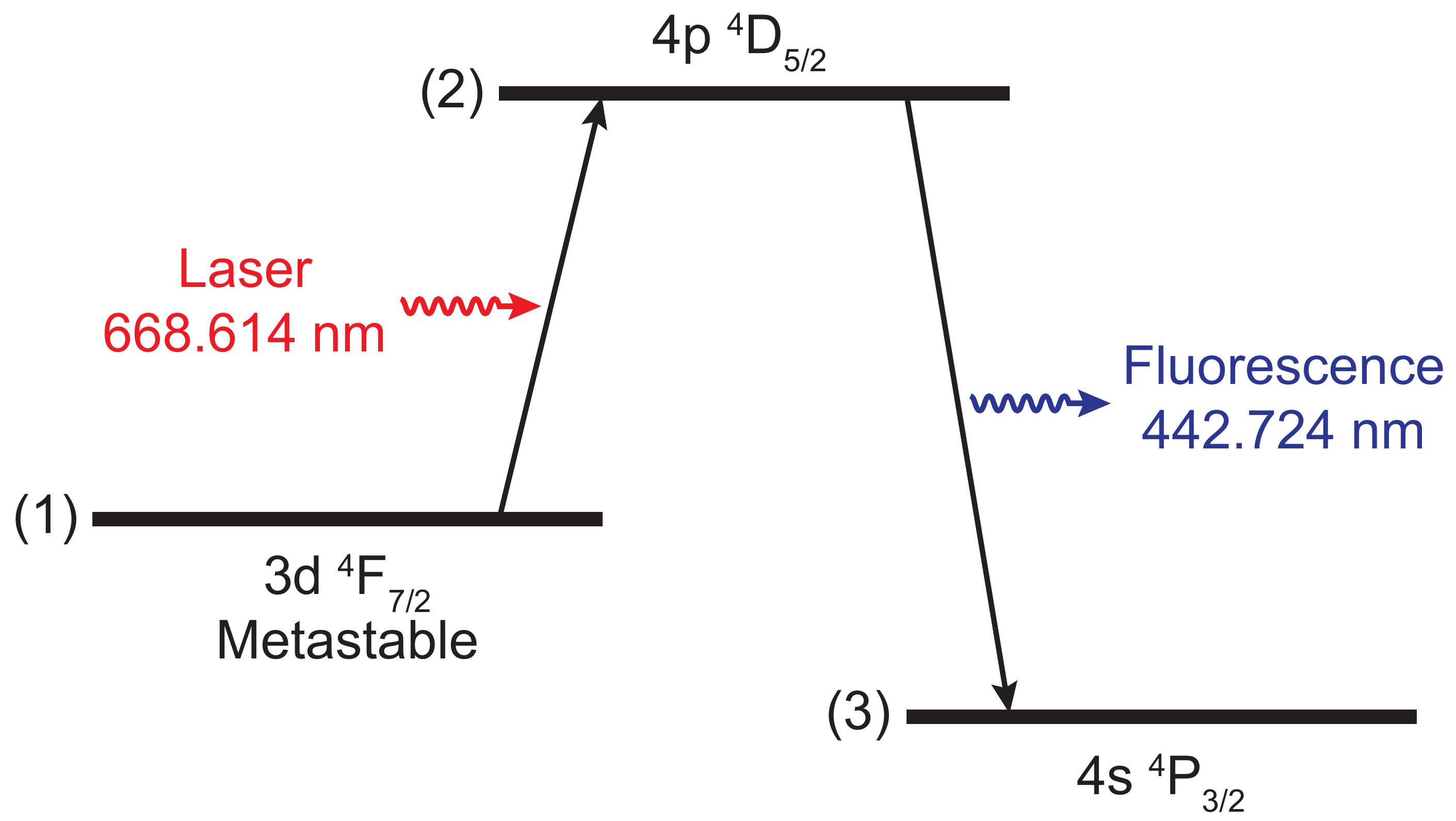}
\caption{Typical energy level diagram for LIF in ArII. This scheme is commonly used, because it provides an adequate LIF signal in a variety of plasmas, the wavelength for the excitation is accessible using single mode diode lasers and many photomultiplier tubes (PMTs) have a high sensitivity between 400 and 500 nm \cite{severn_argon_1998}.}
\label{fig:LIF}
\end{center}
\end{figure}

In order to perform LIF, an electron level transition in ions is stimulated, and the resultant emission of photons due to the decay is detected. Because the natural linewidth tends to be narrow compared to Doppler broadening, the use of a single-frequency laser easily allows for Doppler selection of the ion motion along the direction of the laser beam to the level of 100 $\textup{m/s}$ \cite{skiff_mini-conference_2004}.

\begin{figure}
\begin{center}
\includegraphics[width=2.8in]{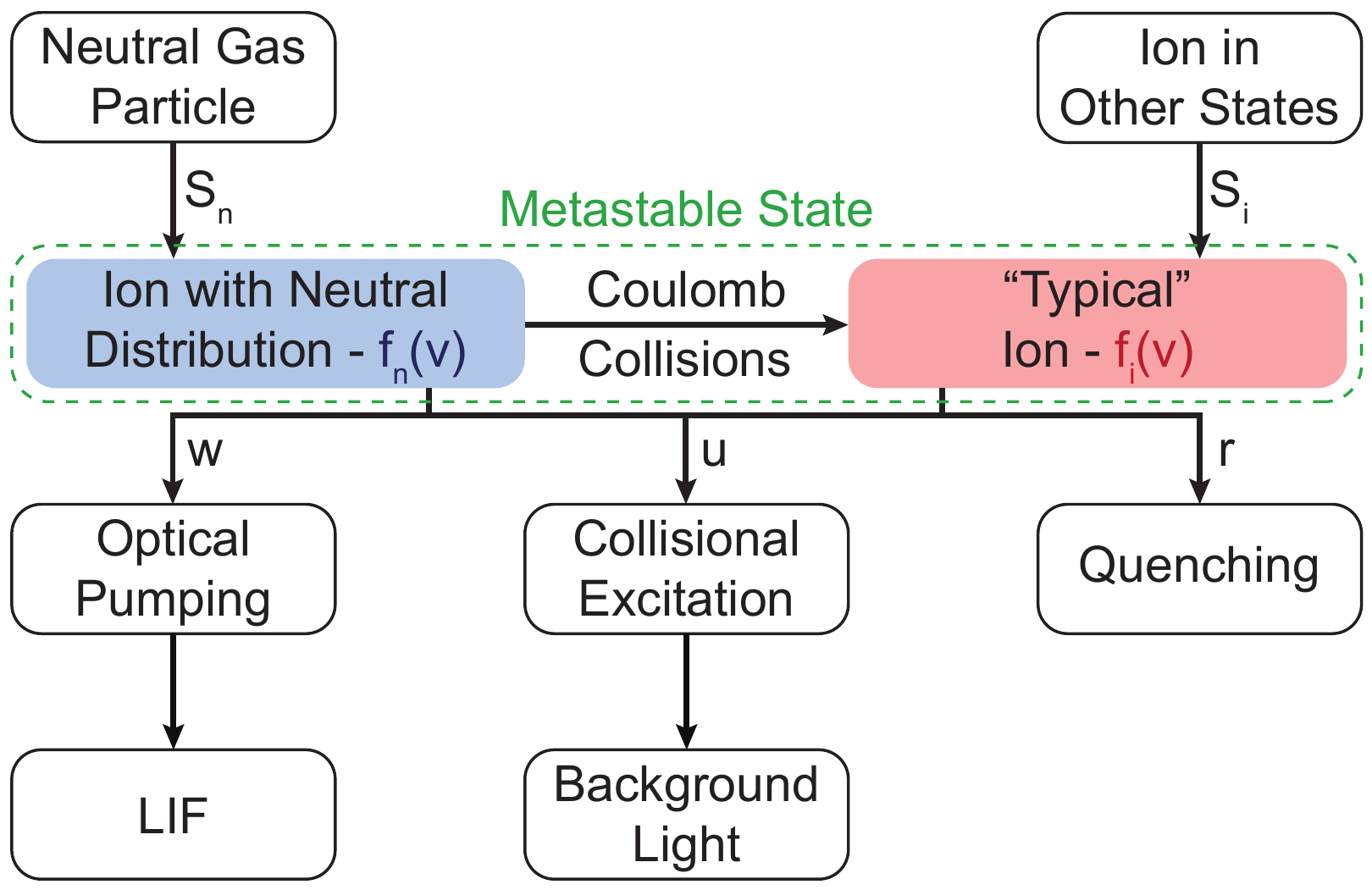}
\caption{Illustration of the production and loss mechanisms for a metastable ion. $S_\textup{n}$ is the metastable birth rate from a neutral particle and $S_\textup{i}$ is birth rate from an ion in other electronic states. $f_\textup{n}(v)$ and $f_\textup{i}(v)$ are the neutral and ion distribution functions respectively.}
\label{fig:dia}
\end{center}
\end{figure}

\begin{figure*}
\begin{center}
\includegraphics[width=6.8in]{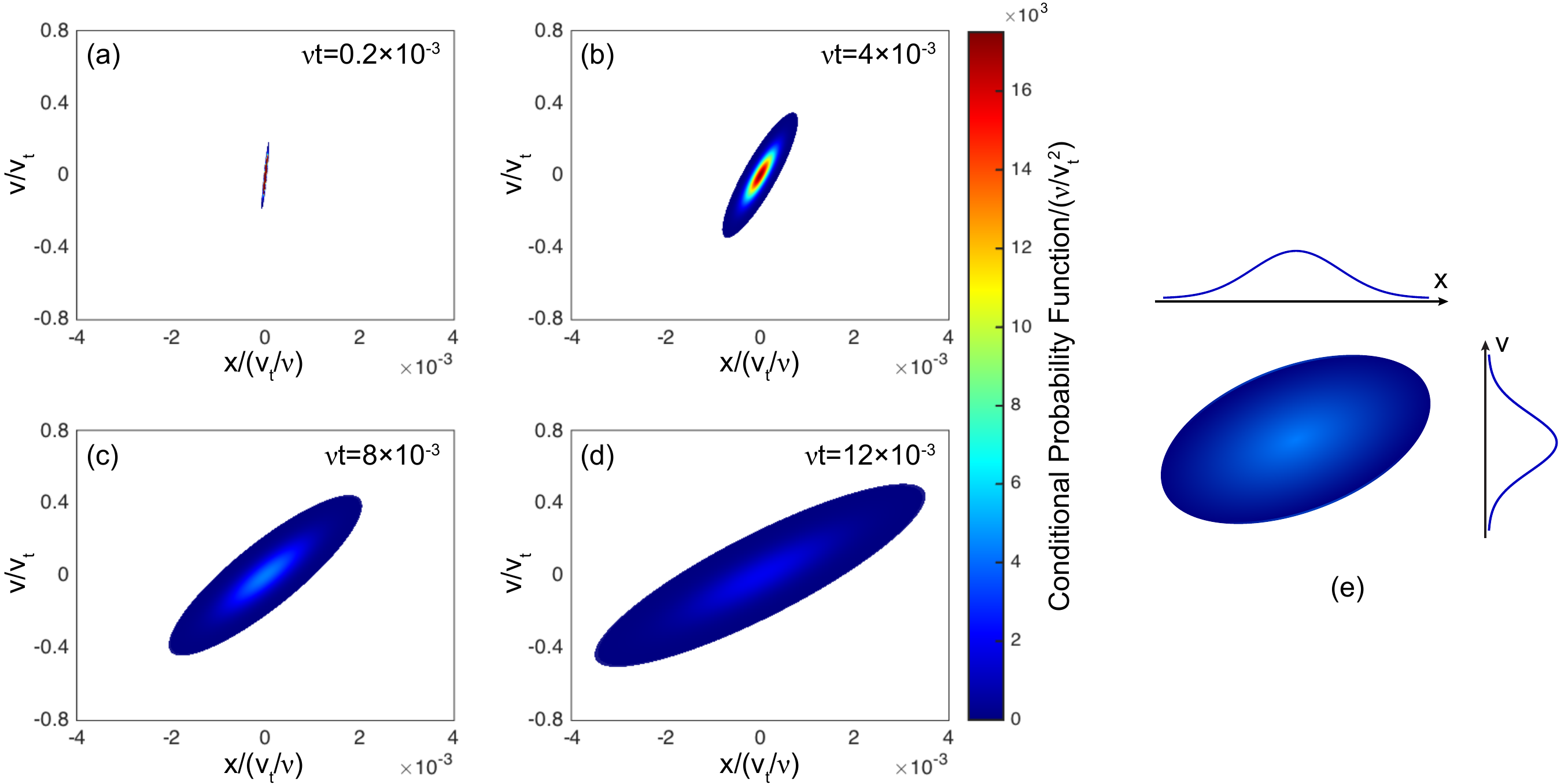}
\caption{(a)--(d) Time evolution of the conditional probability function $P(\mathbf{x},t;\mathbf{x'},t')$ for an ion in phase-space. The probability of finding an ion in the colored phase-space volumes is more than $99\%$. The ion orbit $\mathbf{\tilde{x}}(t)$ is only perturbed by velocity-space diffusion. The thermal velocity of the ions is denoted by $v_\textup{t}$. The initial conditions used in Eq. (\ref{eq:P}) are $x'=0$, $v'=0$. (e) Function $P$ is a 2-D Gaussian function in phase-space.}
\label{fig:P}
\end{center}
\end{figure*}

Laser-induced fluorescence is typically described using a three-level system. A commonly used energy level diagram for ArII \cite{mattingly_measurement_2013} is shown in Fig.~\ref{fig:LIF}. To induce fluorescence, in the rest frame of an ion, a laser is tuned at 668.614 nm to excite electrons in the $\textup{3d }^4\textup{F}_{7/2}$ metastable state to the $\textup{4p }^4\textup{D}_{5/2}^{o}$ state. Fluorescence photons are emitted at 442.724 nm when those electrons decay to the $\textup{4s }^4\textup{P}_{3/2}$ state with a branching ratio of $61.6\%$ \cite{severn_argon_1998}. In the Eulerian approach, one computes the ion distribution function for each energy level. However, in the Lagrangian approach we focus on the probability of a particular ion to be in each level. The rate equations that govern this three-level system are:
\begin{subequations}
\begin{align}
\label{eq:rate1}
\frac{dn_1}{dt}&=-(w+r+u)n_1+A_{21}n_2,\\
\label{eq:rate2}
\frac{dn_2}{dt}&=-A_\textup{T}n_2+(w+u)n_1,\\
\label{eq:rate3}
\frac{dn_3}{dt}&=A_{23}n_2,
\end{align}
\end{subequations}
where $r$ is the metastable state quench rate, $u$ is the electron-collisional excitation rate, $A_{ij}$ is the Einstein coefficient of spontaneous emission, $A_\textup{T}$ is the total spontaneous decay rate of the excited state, and $w$ is the optical pumping rate. The probability that an ion is in level 1 and 2 is denoted by $n_1$ and $n_2$ respectively. The rate of increase of $n_3$ is proportional to the LIF signal. The initial conditions are $n_1=1$ and $n_2=n_3=0$. Stimulated emission can be ignored here because it is smaller than the spontaneous decay rates even in the relatively strong optical pumping regime.

The production and loss mechanisms for a metastable ion are shown schematically in Fig.~\ref{fig:dia}. Metastables are produced from two sources: neutral particles and ions in other electronic states. In the latter case, the metastable ion has a history as a ``typical" ion. In the former case, however, the newly produced metastable ion is representative of the neutral velocity distribution and only becomes ``typical" over time through ion-ion coulomb collisions. Once a metastable is produced, it can be lost primarily through three mechanisms: optical pumping, collisional excitation and quenching. The first contributes to the LIF signal and the second produces background fluorescence light.

\section{1-D Lagrangian Model for LIF}
\label{sec:theo}

Optical pumping is a time-dependent process which typically depends on the ion orbit. The Lagrangian approach achieves large computational advantages by exploiting the separation of the classical dynamics of the ions from the quantum mechanics of the electronic states. First one models the ion orbit, then computes optical pumping, which is a function on the orbit. The total probability distribution of finding a metastable ion at its final position $(x,v)$ in phase-space at time $t$ after it was produced at the initial position $(x',v')$ at time $t'$ can be expressed as
\begin{equation}
\label{eq:dist}
Q (\mathbf{x},t;\mathbf{x'},t')=P(...) \cdot \Psi(P,...),
\end{equation}
where $\mathbf{x}=(x,v)$, $\mathbf{x'}=(x',v')$, $P$ is the conditional probability function that describes the metastable ion orbits, and $\Psi$ represents the quantum state probability. These two functions will be derived in Secs. \ref{subsec:con} and \ref{subsec:the}. 

\subsection{Conditional Probability Function}
\label{subsec:con}

\begin{figure}
\begin{center}
\includegraphics[width=3.2in]{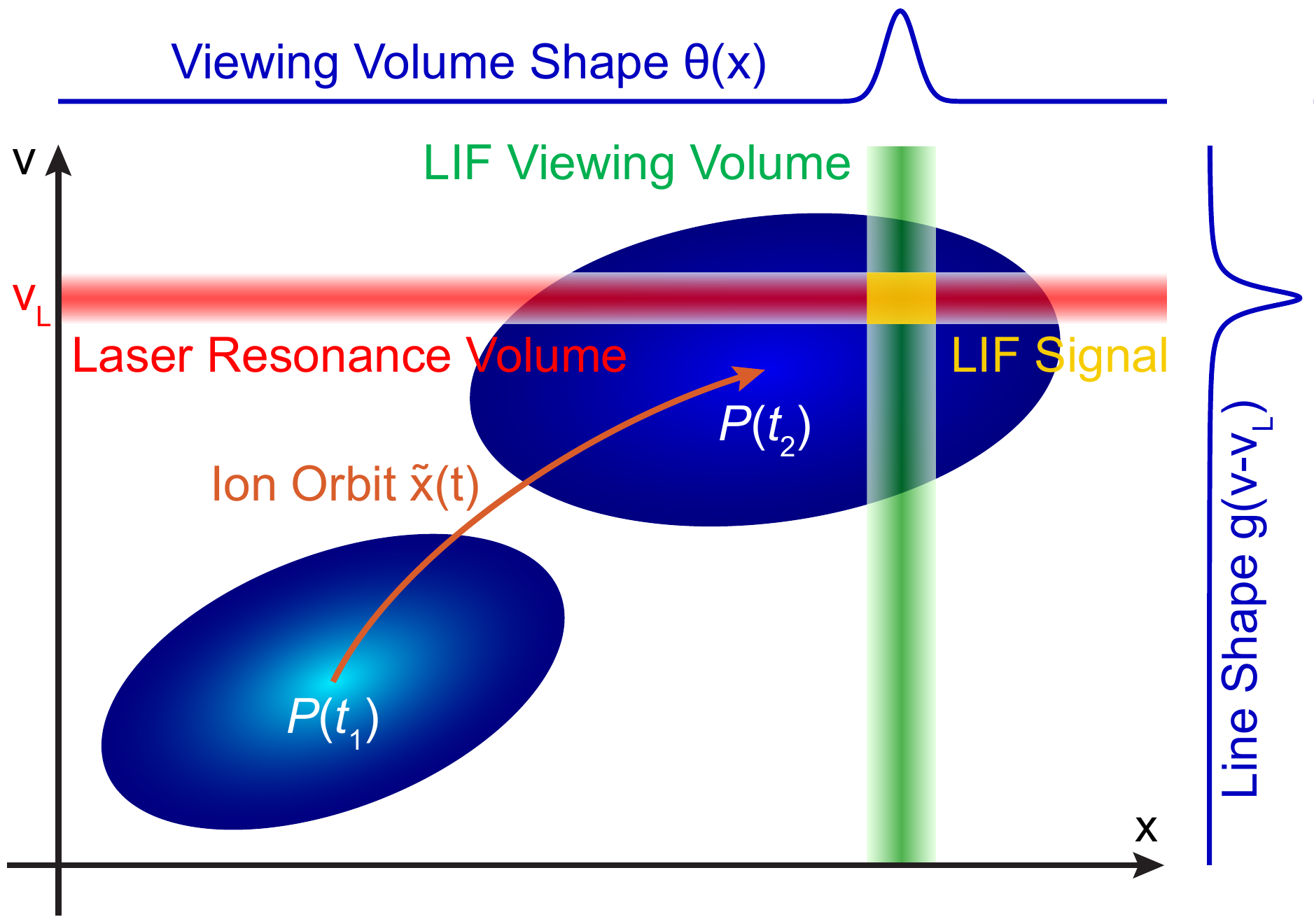}
\caption{The optical pumping process in phase-space. The blue ellipses represent the conditional probability function $P(\mathbf{x},t;\mathbf{x'},t')$ at time $t_1$ and $t_2$. The ion orbit $\mathbf{\tilde{x}}(t)$ without velocity-space diffusion is shown with an orange arrow. The LIF viewing volume is marked in green and the laser resonance volume marked in red. These two volumes are characterized by a normalized window function $\theta(x)$ and natural line shape function $g(v-v_\textup{L})$, respectively. The intersection of these two volumes, which is where the LIF signal is produced, is marked in yellow. }
\label{fig:op}
\end{center}
\end{figure}

The diffusion of particles in velocity-space is a fundamental phenomena in plasma \cite{skiff_plasma_1989, bowles_velocity_1992}. It is important in LIF measurements with a single-frequency laser even at low ion-ion coulomb collision frequency. Without velocity-space diffusion, the Doppler selected metastable ions can be rapidly depleted by the laser beam and the LIF signal correspondingly reduced. To model velocity-space diffusion, a conditional probability function $P$ is introduced. It specifies the probability of finding an ion at the phase-space point $\mathbf{x}$ at time $t$, given that the ion was at point $\mathbf{x'}$ at time $t'$. This function is the Green's function of the kinetic equation for the ions \cite{dougherty_model_1964, chandrasekhar_stochastic_1943}. Adopting a simple 1-D Fokker-Planck model with a constant coulomb collision frequency between ions, the function $P$ is given by
\begin{equation}
\label{eq:P}
P (\mathbf{x},t;\mathbf{x'},t')=p(t-t')\exp \left[-\frac{1}{2}\left ( \mathbf{x}-\mathbf{\tilde{x}} \right ) \mathbf{q} \left(\mathbf{x}-\mathbf{\tilde{x}}\right)^{\mathsf{T}} \right]
\end{equation}
where $\mathbf{\tilde{x}}(t)=(\tilde{x},\tilde{v})$ is the ion orbit in the absence of velocity-space diffusion starting at $\mathbf{\tilde{x}}(0)=\mathbf{x'}$ (see Fig.~\ref{fig:op}). For simplicity, the electric field $\mathbf{E}$ and magnetic field $\mathbf{B}$ are dropped from the kinetic equation. Instead, their contributions are included in the ion orbit $\mathbf{\tilde{x}}(t)$. This treatment provides a first-order approximation. The matrix $\mathbf{q}$ is defined as
\begin{equation}
\label{eq:q}
\mathbf{q}=\begin{pmatrix}
\frac{\nu ^3\left (1+e^{\nu t} \right )}{2\eta \left [ e^{\nu t}(\nu t-2)+\nu t+2 \right ]} &
\frac{\nu ^2\left (1-e^{\nu t} \right )}{2\eta \left [ e^{\nu t}(\nu t-2)+\nu t+2 \right ]} \\[14pt]
\frac{\nu ^2\left (1-e^{\nu t} \right )}{2\eta \left [ e^{\nu t}(\nu t-2)+\nu t+2 \right ]} & 
\frac{\nu \left [ e^{2\nu t}(2\nu t-3)+4e^{\nu t}-1 \right ]}{2\eta \left(e^{\nu t}-1 \right) \left [ e^{\nu t}(\nu t-2)+\nu t+2 \right ]} \nonumber
\end{pmatrix},
\end{equation}
and $p(t)$ is given by
\begin{equation}
\label{eq:p(t)}
p(t)=\frac{\nu ^2 e^{\nu t}}{2 \sqrt{2} \pi \eta }\left \{\left ( e^{\nu t}-1 \right )\left [ \nu t+e^{\nu t}(\nu t-2)+2 \right ] \right \}^{-\frac{1}{2}} \nonumber,
\end{equation}
where $\nu$ is the ion-ion collision frequency, $\eta=\nu T_\textup{i}/m_\textup{i}$, $T_\textup{i}$ is the ion temperature in energy units, and $m_\textup{i}$ is the ion mass. 

\begin{figure}
\begin{center}
\includegraphics[width=3.2in]{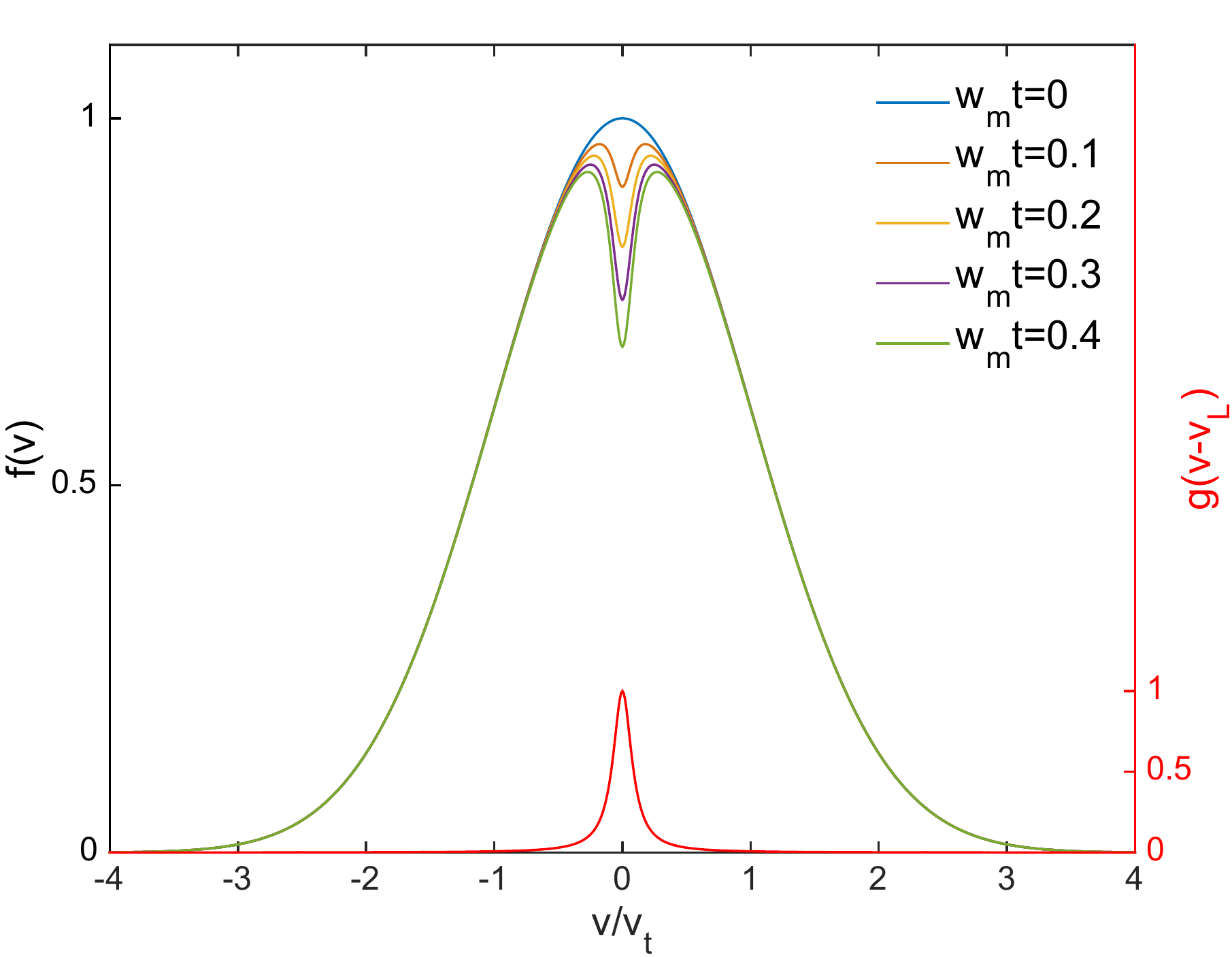}
\caption{Simulated time evolution of the metastable ion distribution function $f(v)$ during optical pumping. The top curve is the metastable ion distribution at thermal equilibrium without optical pumping. The bottom curve in red is the natural line shape function $g(v-v_\textup{L})$ in velocity units. This hole-drilling effect is also simulated using an Eulerian approach, and the results are the same as the Lagrangian approach within the numerical errors.}
\label{fig:op-laser}
\end{center}
\end{figure}

The conditional probability function $P(\mathbf{x},t;\mathbf{x'},t')$ is a 2-D Gaussian function in phase-space. In other words, the projection of $P$ on either dimension, $x$ or $v$, is a 1-D Gaussian function, as shown in Fig.~\ref{fig:P}(e). $P$ is shown as it evolves in time in Figs.~\ref{fig:P}(a)--(d) where the probability of finding a metastable ion is color coded. Probability is higher near the center and lower near the edge. $P$ starts as a 2-D delta function where the metastable ion is produced, and then gradually relaxes to a 2-D Gaussian function over time due to velocity-space diffusion. The elliptical shape tilts towards the $x$ axis as it evolves since higher speeds travel further than the lower speeds in the frame of the metastable ion orbit $\mathbf{\tilde{x}}(t)$.

\subsection{Optical Pumping}
\label{subsec:opt}

Optical pumping is a process in which absorption of light produces a change in the energy level populations. For metastable ions with velocity $v$ and single-frequency laser intensity $I$, the optical pumping rate for a given transition can be expressed as
\begin{equation}
\label{eq:w}
w(v,t)=\frac{BI(t)}{c\gamma }g(v-v_\textup{L}),
\end{equation}
where $B$ is the Einstein coefficient of induced absorption, $\gamma$ is the half linewidth of the transition in frequency units, $g$ is a dimensionless Lorentzian function that represents the natural line shape of the transition in velocity space, and $v_\textup{L}$ is the velocity that an ion must have to Doppler shift the laser light into resonance with the transition. The line shape $g$ is
\begin{equation}
\label{eq:g}
g(v-v_\textup{L})=\frac{1}{\pi}\cdot \frac{\left ( \gamma \lambda _\textup{L} \right )^{2}}{\left ( v-v_\textup{L} \right )^2+\left ( \gamma \lambda _\textup{L} \right )^2} ,
\end{equation}
where $\lambda_\textup{L}$ is the laser wavelength. 

A sketch of the optical pumping process is shown in Fig.~\ref{fig:op}. The intersection of the LIF viewing volume and the laser resonance volume, which is where the LIF signal is produced, selects the ions with specific $x$ and $v$ in phase-space. The shapes of these two volumes are characterized by a normalized window function $\theta(x)$ and natural line shape function $g(v-v_\textup{L})$. The total optical pumping rate for a metastable ion at time $t$ is calculated by averaging over the conditional probability function:
\begin{equation}
\label{eq:Wn}
W(t;\mathbf{x'},t' )=\int w\left (v,t \right )P\left ( \mathbf{x},t;\mathbf{x'},t' \right )d\mathbf{x}.
\end{equation}
The integral in Eq. (\ref{eq:Wn}) is a convolution of a Lorentzian and a Gaussian function, i.e., a Voigt function, which can be evaluated analytically through a continued fraction expansion of the plasma dispersion function \cite{armstrong_spectrum_1967,mccabe_continued_1984}.

The simulated evolution of the metastable ion distribution function along with the natural line shape function are shown in Fig.~\ref{fig:op-laser}. The maximum optical pumping rate in Eq. \ref{eq:w} is denoted by $w_\textup{m}=BI_0/c\gamma \pi$. Fig.~\ref{fig:op-laser} shows that the velocity resolution of LIF can be affected by optical pumping through drilling a hole in the distribution of available metastables. This hole-drilling effect was observed experimentally by Klimcak and Camparo \cite{klimcak_optical-pumping_1984}.


\subsection{Theoretical Model}
\label{subsec:the}


The lifetime of the upper energy state (level 2 shown in Fig.~\ref{fig:LIF}) is typically short compared to the ion dynamical scales. Thus, the probability that an ion is in the upper state is simply proportional to the probability that it is in the metastable state:
\begin{equation}
\label{eq:N2N1}
n_2=\frac{w+u}{A_\textup{T}}n_1.
\end{equation}
Inserting this into Eq. (\ref{eq:rate1}) gives the time dependent probability of an ion to remain in the metastable state. The solution for this time dependent probability is $\Psi$:
\begin{eqnarray}
\label{eq:psi}
\Psi (t;\mathbf{x'},t')=&&\exp \Bigg\{ - \bigg[ \bigg(1-\frac{A_{21}}{A_\textup{T}} \bigg ) \int_{t'}^{t}W \left ( t''; \mathbf{x'},t' \right )dt'' \nonumber \\
&&+\bigg(1-\frac{A_{21}}{A_\textup{T}} \bigg )u\left ( t-t' \right )+r\left ( t-t' \right ) \bigg ] \Bigg \},
\end{eqnarray}
where the initial value $\Psi (t';\mathbf{x'},t')$ is unity. Therefore the total probability distribution function $Q$ can be constructed by combining Eqs. (\ref{eq:P}) and (\ref{eq:psi}):
\begin{equation}
\label{eq:Q}
Q (\mathbf{x},t;\mathbf{x'},t')=P (\mathbf{x},t;\mathbf{x'},t') \Psi (t;\mathbf{x'},t').
\end{equation}



In this three-level system, every observable fluorescence photon comes from an optically pumped ion that transitions from level 2 to level 3. According to Eq. (\ref{eq:rate3}), the contribution of a metastable ion produced at the initial phase-space point $\mathbf{x'}$ at time $t'$ to the fluorescence signal (in photons/s) in the viewing volume is
\begin{equation}
\label{eq:signal}
N (t;\mathbf{x'},t')=\frac{A_{23}}{A_\textup{T}}\iint\left [w(v,t)+u \right ]Q(\mathbf{x},t;\mathbf{x'},t')\theta (x)d\mathbf{x}.
\end{equation}
Finally, the total signal can be calculated by summing over all the initial conditions: 
\begin{equation}
\label{eq:finalsignal}
N (t)=S_\textup{n}\iint\limits_{\mathbf{x'}}\int_{-\infty}^{t}N ( t;\mathbf{x'},t')f_0(v',t')d\mathbf{x'}dt'.
\end{equation}
where $S_\textup{n}f_0(v',t')$ is the metastable birth rate from neutrals.


\section{Experimental Setup}
\label{sec:exp}

\begin{figure}
\begin{center}
\includegraphics[width=3.4in]{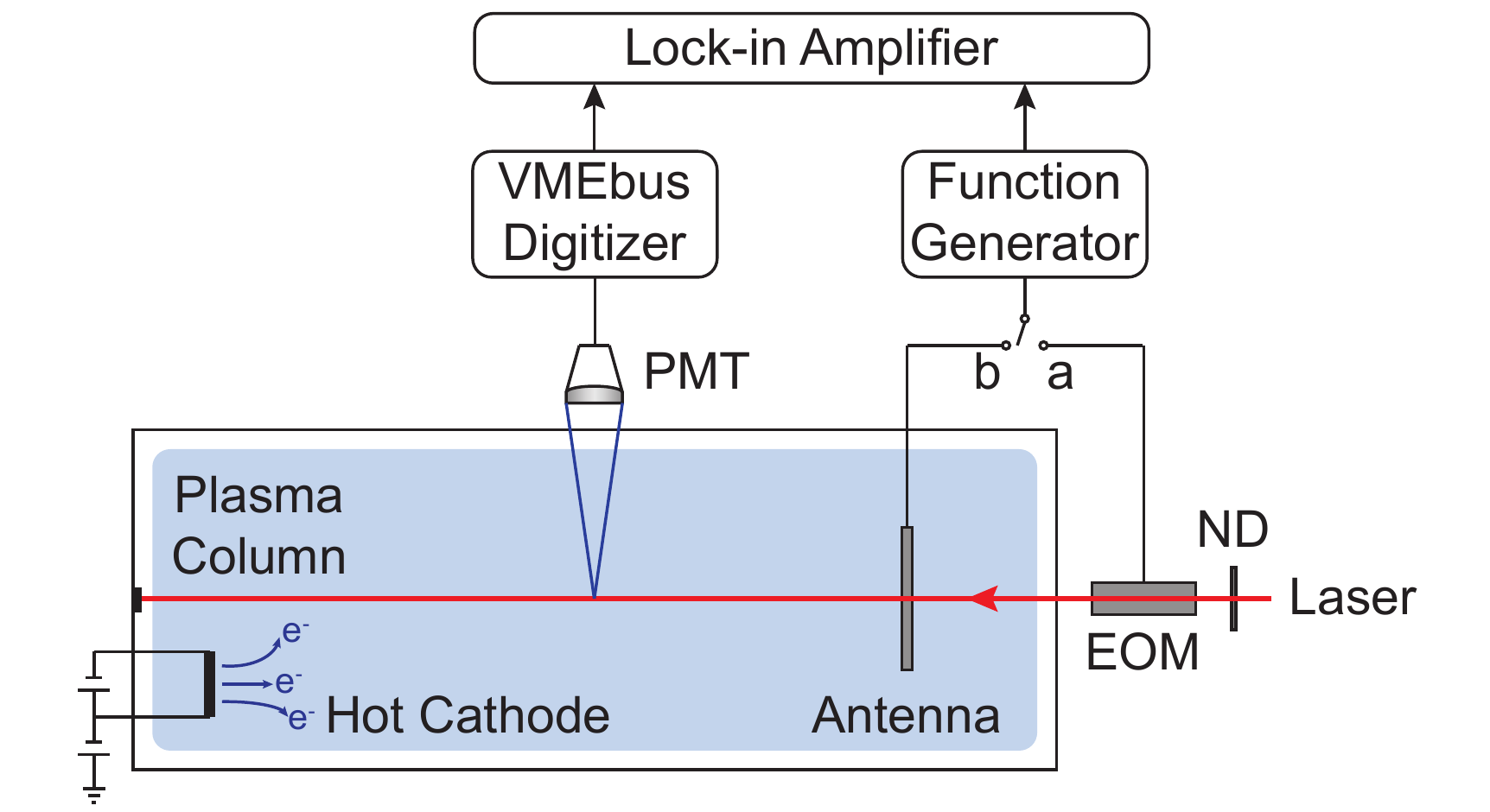}
\caption{Schematic of the experimental setup. The LIF diagnostic equipment (not drawn to scale) includes a set of removable neutral-density (ND) filters, an electro-optic modulator (EOM), a mesh antenna of 81 $\%$ open area, a 16-channel photomultiplier tube (PMT), a Versa Module Europa bus (VMEbus) board, and a lock-in amplifier. In the IVDF measurement, the EOM is connected to the function generator to chop the laser beam. In the wave measurement, the mesh antenna is connected to the function generator instead to excite ion acoustic waves in the plasma. The antenna is 5 cm away from the LIF viewing volume.}
\label{fig:exp}
\end{center}
\end{figure}

Experimental tests are carried out in an Argon multidipole plasma to validate our new model for LIF. The plasma is produced through impact ionization by primary electrons emitted from a hot cathode biased at $-75$ V, resulting in an emission current of 72.9 mA. The multidipole magnetic field is provided by 16 rows of magnets with alternating poles covering all inside walls of the chamber. On the surface of the magnets, the field is 1000 G. The field quickly diminishes to less than 2 G in the measurement region. Neutral pressure is regulated at $5.6 \times 10^{-5}$ Torr by a mass flow controller. More details of this multidipole chamber are found in Ref. \onlinecite{hood_ion_2016}.

The LIF scheme used in the experiment is described in Sec. \ref{sec:LIF} and Fig. \ref{fig:LIF}, which is accomplished by a single mode tunable diode laser (Toptica TA 100) with a narrow bandwidth of only 1 MHz. To adjust the beam intensity without affecting the laser operation, neutral-density (ND) filters are inserted in the beam path between the laser and the electro-optic modulator (EOM). 

The experimental setup is shown in Fig.  \ref{fig:exp}. For the purpose of the IVDF measurement, the laser beam is chopped by the EOM at 62.5 kHz. This chopping signal is then sent to the lock-in amplifier as a reference. To measure the perturbation of IVDFs, a mesh antenna is connected in the circuit instead of the EOM. A square waveform of $V_\textup{p}=1\textup{V} \sim 5\textup{V}$ and $f=10$ kHz is applied on the antenna to excite ion acoustic waves in the plasma.

Using LIF, we find that the ion temperature $T_\textup{i}=0.03$ eV along the direction of the laser beam in the center of the chamber. The other plasma parameters are measured by a disc-shaped Langmuir probe with electron density $n_\textup{e}=2.52 \times 10^9$ $\textup{cm}^{-3}$, electron temperature $T_\textup{e}=2.94$ eV, and plasma potential $V_\textup{p}=2.45$ V. This gives an ion sound speed $C_\textup{s}$ of approximately $2.6 \times 10^5$ $\textup{cm}$  $\textup{s}^{-1}$.

\section{Simulation and Experimental Results}
\label{sec:results}

\begin{figure}
\begin{center}
\includegraphics[width=3.2in]{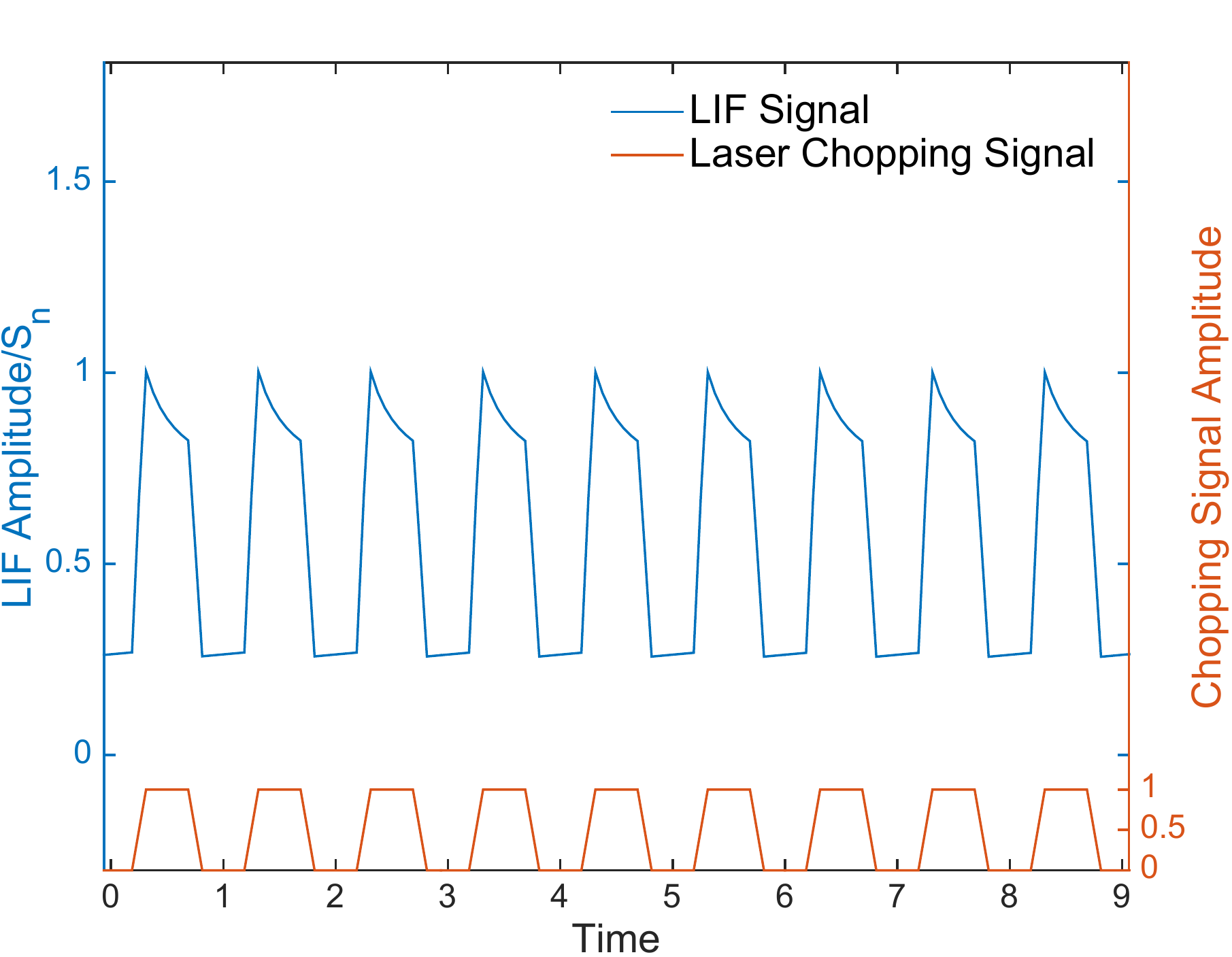}
\caption{A time series of the LIF signal from the simulation. Time is normalized by the laser chopping frequency. The normalized optical pumping rate $w_\textup{m}/S_\textup{n}=21.2$ and the collisional excitation rate $u/S_\textup{n}=0.17$.}
\label{fig:timeseries}
\end{center}
\end{figure}



Based on the above LIF model, a numerical simulation is performed to study how laser intensity, ion-ion coulomb collisions, and metastable quenching affect the LIF measurements. All the simulation results in the rest of the paper are obtained through analyzing the time series of the simulated LIF signal, which can be numerically calculated using Eq. (\ref{eq:finalsignal}).

A typical simulated LIF signal is shown in Fig.~\ref{fig:timeseries}. The two sources of fluorescence are optical pumping and collisional excitation, of which the latter produces background noise. When the laser pulse starts, because of the initial large amount of  metastable ions, the signal suddenly jumps to a high level, and then decays nearly exponentially due to the depletion of these ions by optical pumping. When the laser pulse ends, the background level of fluorescence starts to recover gradually through the birth of new metastable ions.



The basic effects of optical pumping are the same independent of the origin of a metastable ion. For simplicity, the ions and neutrals are assumed to have the same temperature in the results shown below except in Fig.~\ref{fig:f0qu}. The results from Secs. \ref{IVDF} and \ref{pert} apply both to metastables produced from neutral particles and from ions in other electronic states.

\subsection{Ion Velocity Distribution Function}
\label{IVDF}

\begin{figure}
\begin{center}
\includegraphics[width=3.2in]{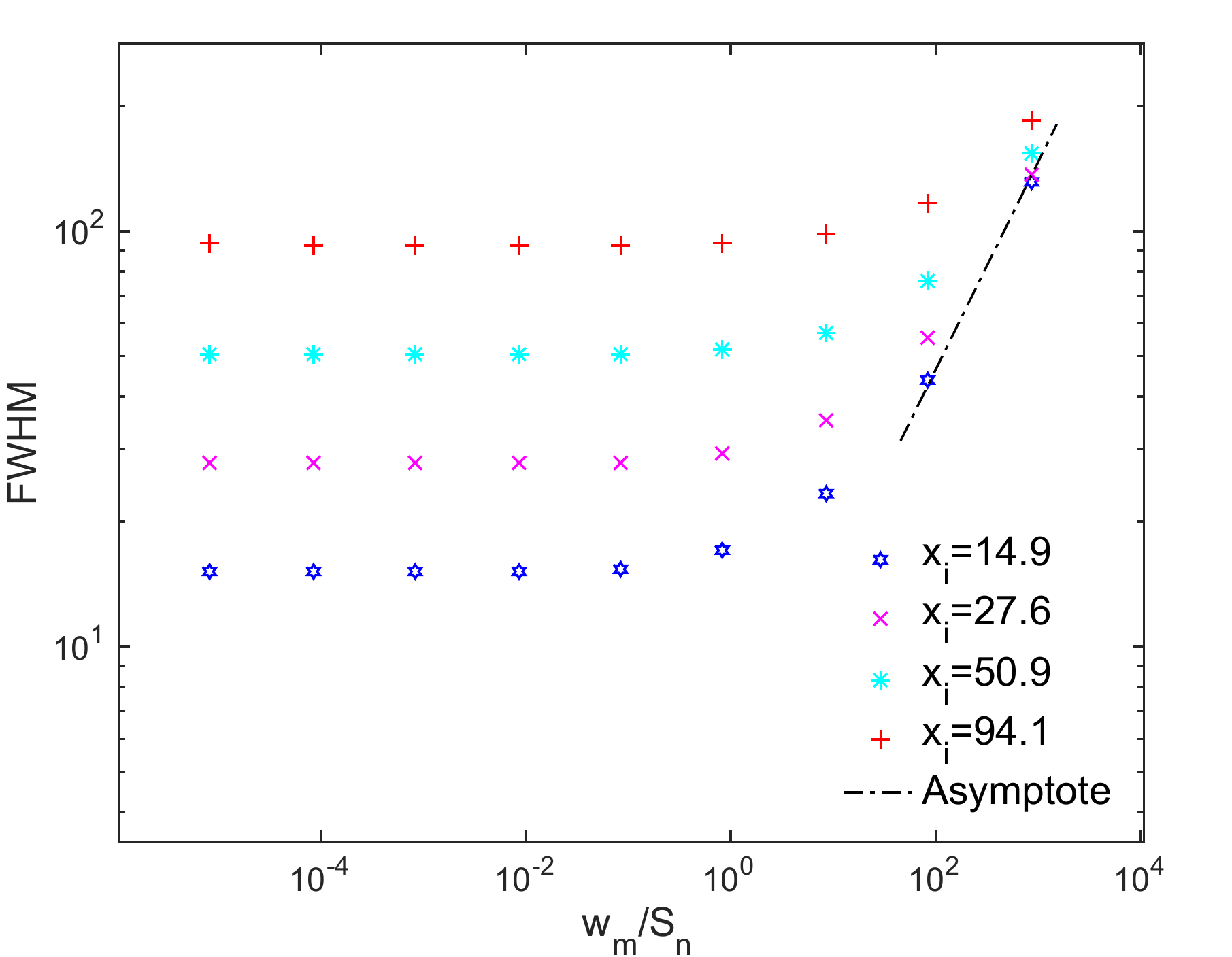}
\caption{Simulated FWHM of the IVDF as a function of optical pumping rate $w_\textup{m}$ for various ion temperatures. $x_\textup{i}$ is the actual full width of the ion velocity distribution. All FWHM values are normalized by the natural linewidth. Optical pumping broadening starts to occur at $w_\textup{m}/S_\textup{n}\sim 1$. At high optical pumping rate, the measured FWHM values are given by Eq. (\ref{eq:FWHM2}), and independent of the ion temperatures. The slope of the asymptote is $1/2$, suggesting that the measured FWHM is proportional to $\sqrt{w_\textup{m}}$ for high optical pumping rates.}
\label{fig:FWHM-in}
\end{center}
\end{figure}

The IVDF is obtained by scanning the laser wavelength \cite{mattingly_measurement_2013}. Ideally, the full width at half maximum (FWHM) of the IVDF for a Maxwellian plasma is given by \cite{goeckner_laserinduced_1989}
\begin{equation}
\label{eq:FWHM}
\Delta _{\textup{FWHM}}=2\sqrt{ \frac{2\ln{2} T_\textup{i}}{m_\textup{i}} }.
\end{equation}
At large laser intensity, optical pumping will deplete all the metastable ion population and cause a broadening in the measured IVDF \cite{klimcak_optical-pumping_1984}. When the laser intensity is sufficiently high, even the laser photons in the wings of the Lorentzian profile can deplete the metastable population. The FWHM is then given by 
\begin{equation}
\label{eq:FWHM2}
\Delta _{\textup{FWHM}}\simeq  2\gamma \lambda_\textup{L}\sqrt{\frac{w_\textup{m}}{r+u}}\propto \sqrt{I_0}.
\end{equation}

The simulated FWHM of the IVDF as a function of optical pumping rate is shown in Fig.~\ref{fig:FWHM-in}. If the optical pumping rate is increased above a certain level, optical pumping broadening occurs and Eq. (\ref{eq:FWHM}) will not provide a correct ion temperature. However, the FWHM has little variation below that level.  

\begin{figure}[t]
\begin{center}
\includegraphics[width=3.2in]{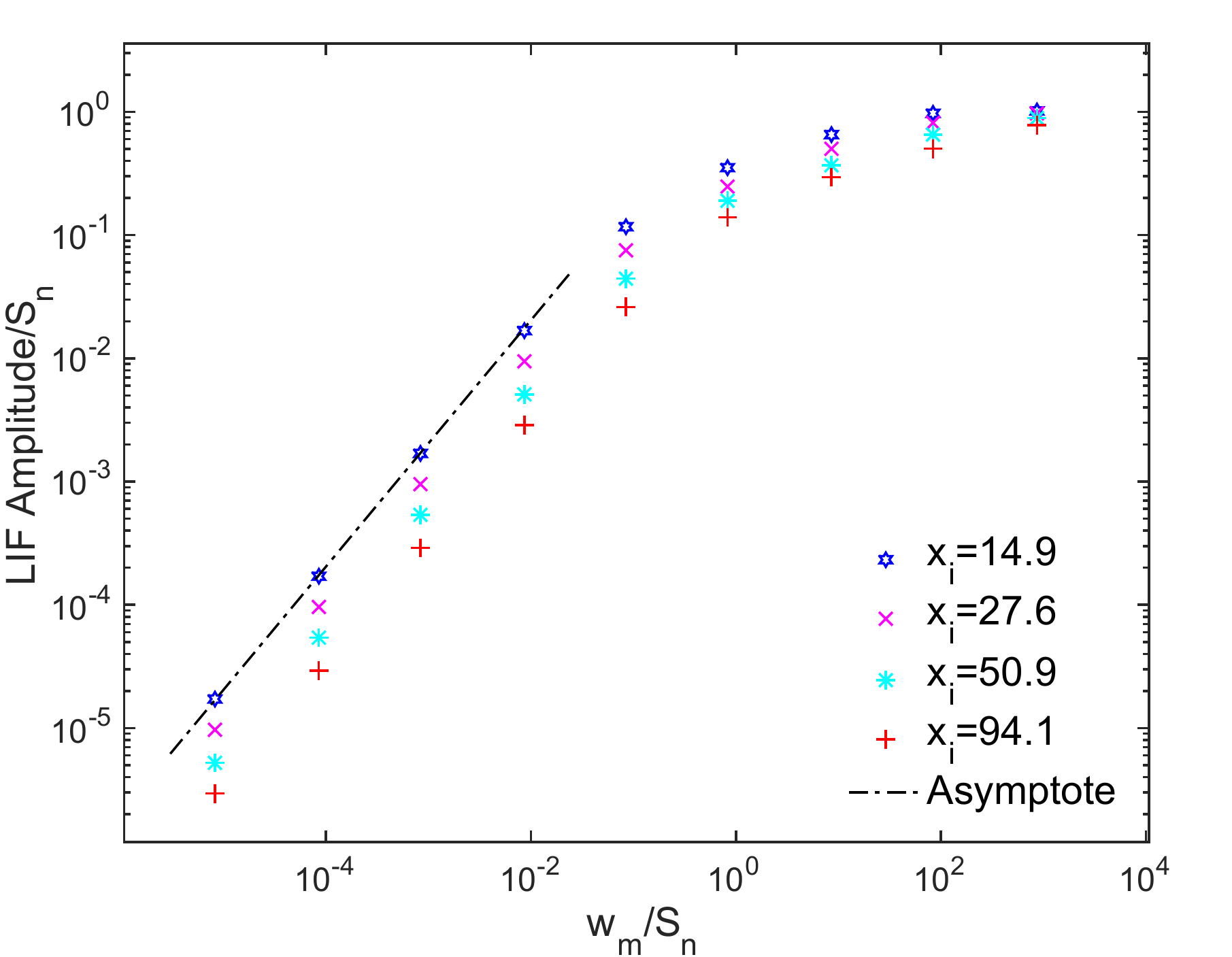}
\caption{Simulated LIF signal amplitude at $v_\textup{L}=0$ as a function of optical pumping rate $w_\textup{m}$ for various ion temperatures. When $w_\textup{m}/S_\textup{n}<1$, the asymptote has a slope of $1$, suggesting that the signal is proportional to the optical pumping rate. Above this level, the signal strength reaches a limit. The LIF signal is larger for ions with a lower temperature, because the distribution function of these ions is less spread out, resulting in more ions in the laser resonance volume around $v_\textup{L}=0$ for a given natural linewidth.}
\label{fig:inten}
\end{center}
\end{figure}

\begin{figure}[t!]
\begin{center}
\includegraphics[width=3.2in]{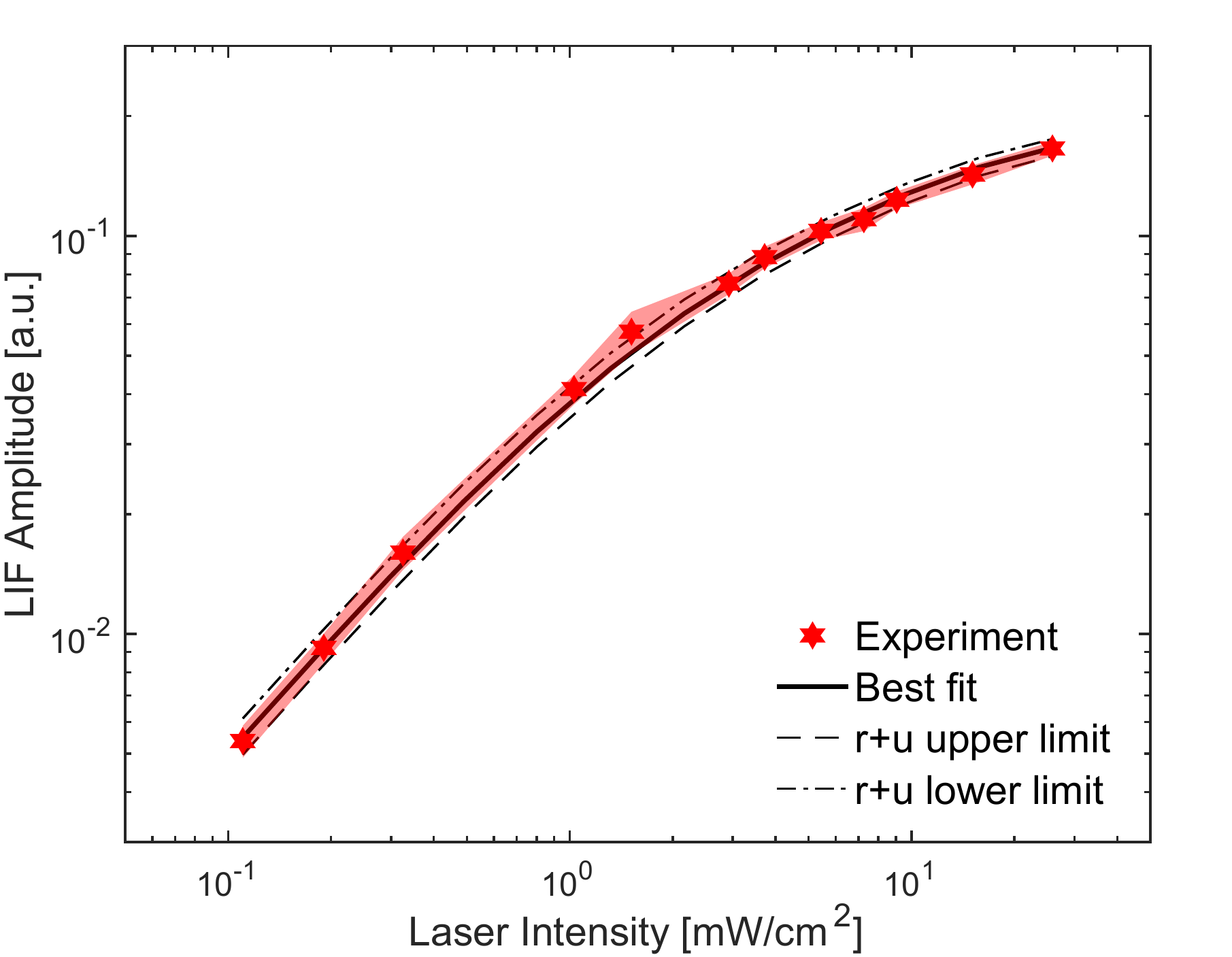}
\caption{Comparison between the simulated and measured LIF signal amplitude at $v_\textup{L}=0$ as a function of the laser intensity. The parameters used in the simulation are $A_{21}=1.07 \times 10^7$ $\textup{s}^{-1}$, $A_{\textup{T}}=1.33 \times 10^8$ $\textup{s}^{-1}$, $B=1.44 \times 10^{20}$ $\textup{J}^{-1} \textup{m}^{3} \textup{s}^{-2}$, $\gamma=10.6$ GHz \cite{whaling_argon_1995}, $\nu=400$ $\textup{s}^{-1}$. The uncertainty of measurements ($\pm \sigma$) is shown in red shade. From the best fit, $r + u$ is estimated to be $(1.68 \pm 0.18) \times 10^4$ $\textup{s}^{-1}$.}
\label{fig:inten-expu}
\end{center}
\end{figure}

This effect can also be seen from another aspect. The simulated LIF signal amplitude at $v_\textup{L}=0$ as a function of optical pumping rate is shown in Fig.~\ref{fig:inten}. The signal scales proportionally to the optical pumping rate when the rate is low. As this rate increases, the signal strength reaches an asymptotic limit, because ultimately the LIF signal is limited by the total number of metastable ions. These results are consistent with Figs. 3--4 in Ref. \onlinecite{goeckner_laserinduced_1989}, where optical pumping broadening was demonstrated by Goeckner \textit{et al}. using an Eulerian approach. The measured LIF signal amplitude at $v_\textup{L}=0$ as a function of the laser intensity is compared with the simulation in Fig. \ref{fig:inten-expu}. From the best fit, the addition of the quench rate and the collisional excitation rate $r+u$ is found to be $(1.68 \pm 0.18) \times 10^4$ $\textup{s}^{-1}$. Since these time scales are difficult to measure in the experiment, estimating their values can be a useful application of the Lagrangian model.

Broadening of the IVDF can be avoided by reducing the laser intensity; however, this will also diminish the fluorescence signal. The balance of a strong signal and minimal broadening determines the optimum laser intensity.

\subsection{Perturbation of IVDF with Electrostatic Waves}
\label{pert}

Electrostatic waves that propagate in a plasma can perturb the IVDF. The ion orbit $\mathbf{\tilde{x}}(t)$ used in the simulation is
\begin{equation}
\label{eq:orbit}
\dot{v}=-\nu v+\frac{Ee}{m_\textup{i}}\sin\left ( kx-\omega t \right ) ,
\end{equation}
where the first term on the right side represents drag on the ion, and $E$ is the wave amplitude.

\begin{figure}
\begin{center}
\includegraphics[width=3.2in]{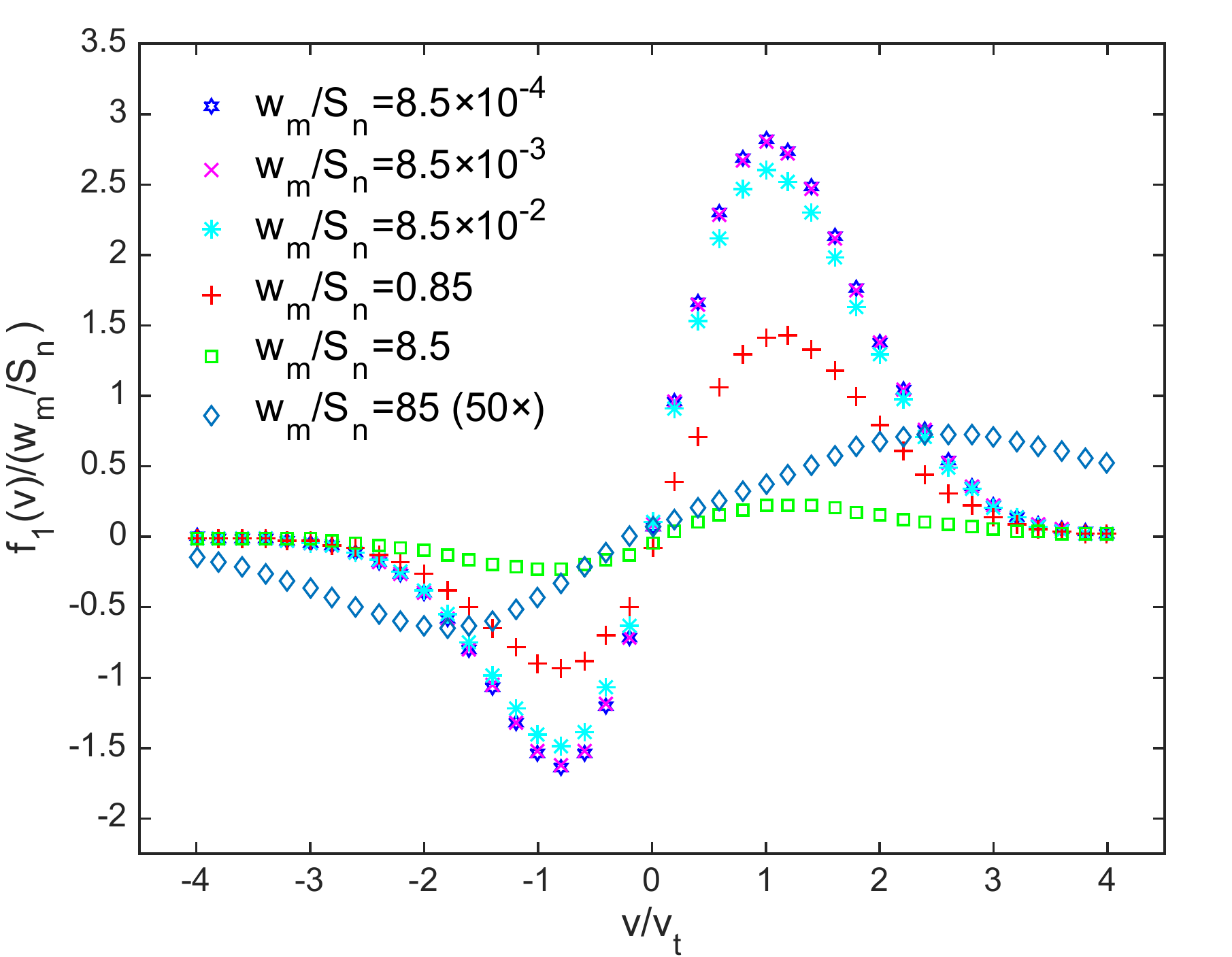}
\caption{Simulated $f_1(v)$ for various optical pumping rates. The $f_1(v)$ in the plot is already normalized by $f_0(v)|_{v=0}$ measured at $w_\textup{m}/S_\textup{n}=8.5\times 10^{-4}$. In the simulation, the phase velocity $v_{\textup{p}}/v_{\textup{t}}=3.8$. The curve for $w_\textup{m}/S_\textup{n}=85$ is magnified by 50 times to scale. At low optical pumping rates, the shapes of $f_1(v)$ almost remain the same. However, continuing to increase the optical pumping rate causes a broadening and reduction in $f_1(v)$.}
\label{fig:f1in}
\end{center}
\end{figure}

\begin{figure}
\begin{center}
\includegraphics[width=3.2in]{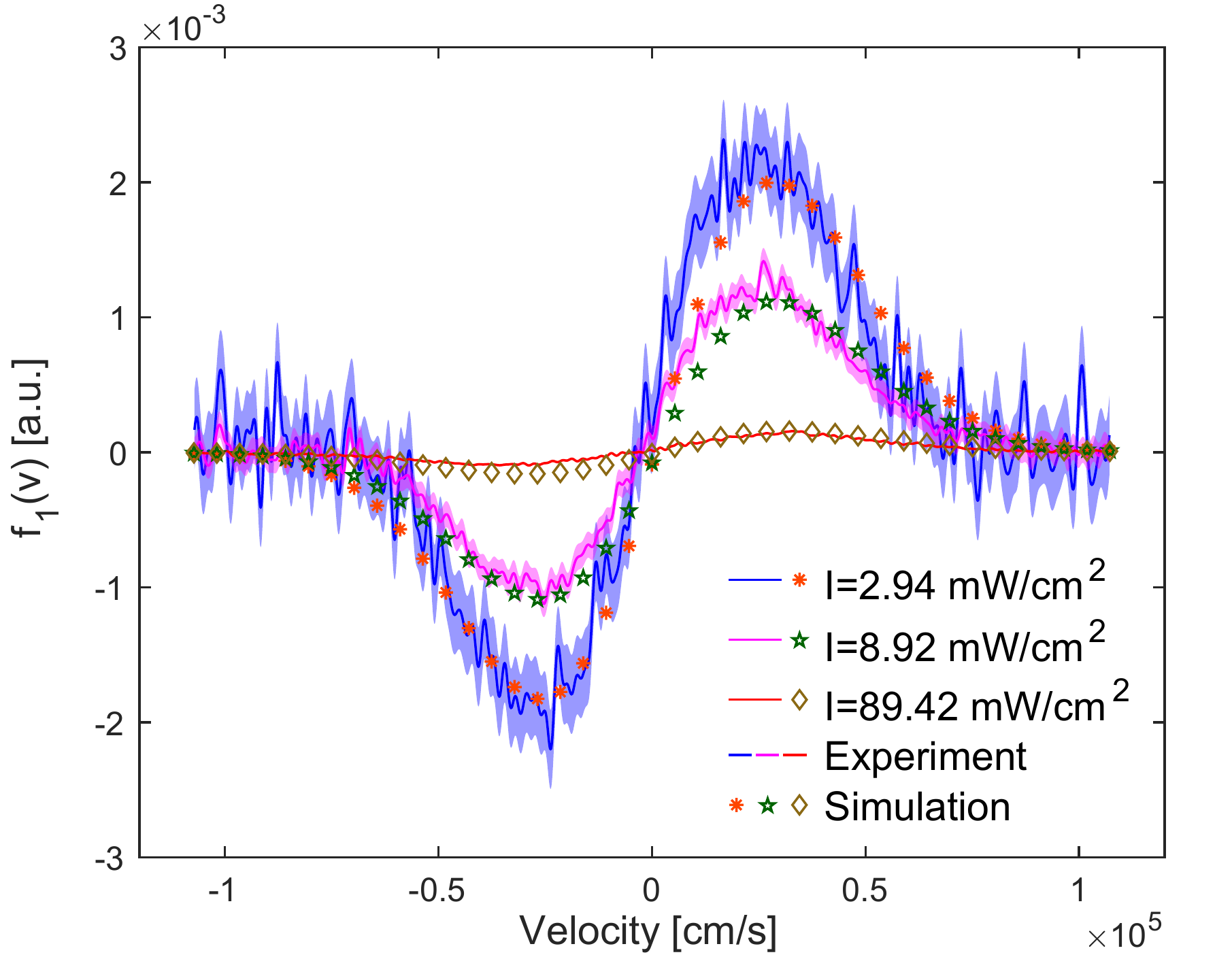}
\caption{Comparison between the simulated and measured $f_1(v)$ for various laser intensities. The curves of $f_1(v)$ are normalized by the laser intensities $I$. The parameters used in the simulation are the same as the ones shown in Fig. \ref{fig:inten-expu}. The uncertainties of measurements ($\pm 3 \sigma$) are shown in colored shades.}
\label{fig:f1-in-exp}
\end{center}
\end{figure}

In a collisional plasma, the first-order perturbation of the IVDF $f_1(v)$ is a complex function. It tends to be in phase with the wave potential when the collisions are weak. For demonstration purposes, only the real part of $f_1(v)$ is presented here. The simulated $f_1(v)$ for various optical pumping rates is shown in Fig.~\ref{fig:f1in}. At low optical pumping rates, the shapes of $f_1(v)$ remain undistorted. However, a broadening and reduction in $f_1(v)$ occur as the optical pumping rate increases. The measured $f_1(v)$ for various laser intensities is compared with the simulation in Fig. \ref{fig:f1-in-exp}.

\begin{figure}
\begin{center}
\includegraphics[width=3.2in]{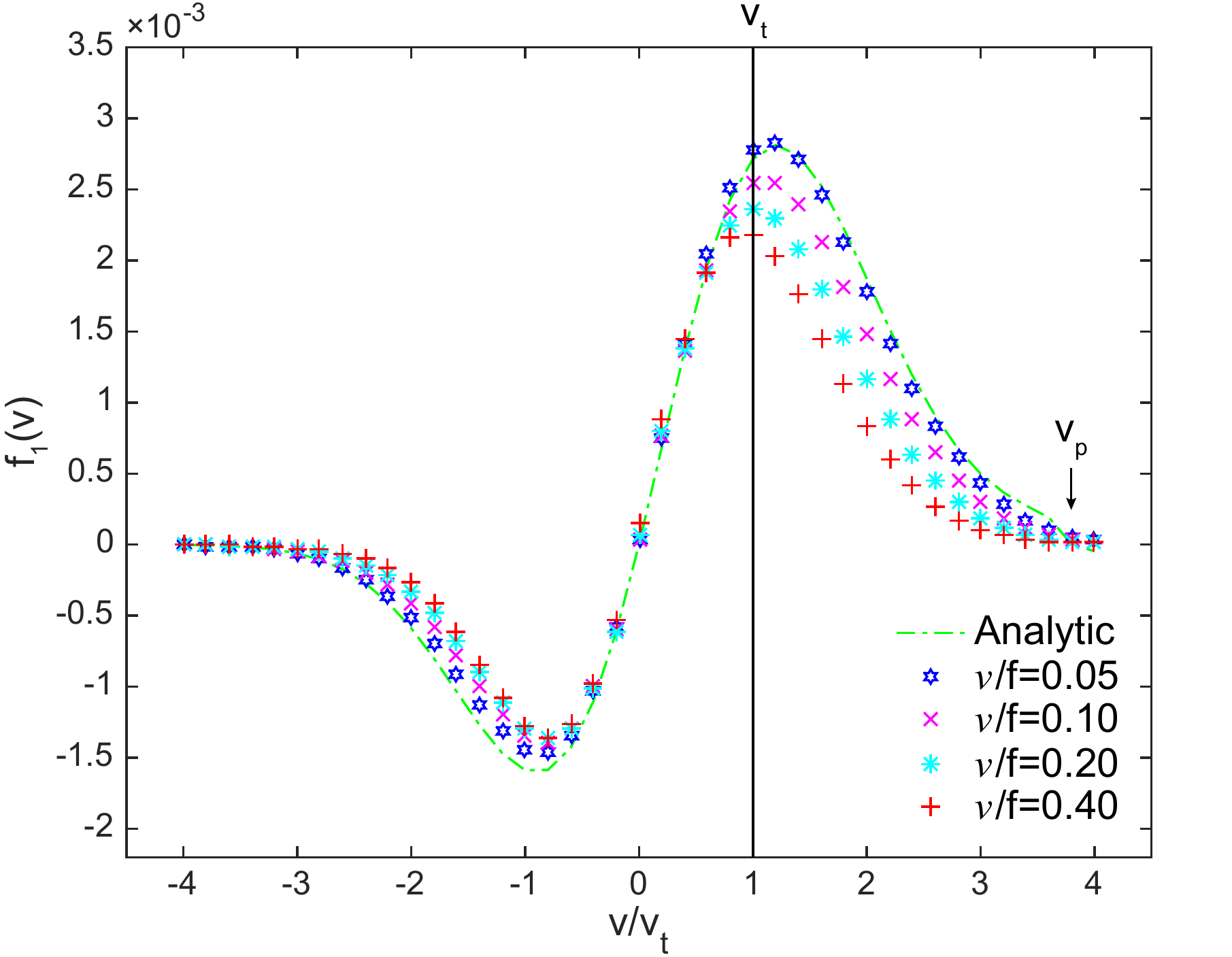}
\caption{Simulated $f_1(v)$ for various ion-ion collision frequencies at $w_\textup{m}/S_\textup{n}=8.5\times10^{-4}$. The collision frequency is normalized by the wave frequency. The $f_1(v)$ in the plot is normalized by $f_0(v)|_{v=0}$ measured at collision frequency $\nu /f=0.05$. The analytic solution of the Vlasov equation is also shown for comparison.}
\label{fig:f1col}
\end{center}
\end{figure}

Fig.~\ref{fig:f1col} shows the simulated $f_1(v)$ for various ion-ion collision frequencies along with the analytic solution of the Vlasov equation for comparison. At a low collision frequency, the measurement agrees well with the Vlasov solution. As the collision frequency increases, the systematic averaging effect makes the measurements more symmetric and behave as if the ion response is local. 

\subsection{Lifetime of Metastable Ions}

As mentioned earlier, metastables can be produced from neutral particles or ions in other electronic states. In the former situation, at the time when the metastable ions are produced, they represent the neutral distribution and only become ``typical" ions at a later time. Since LIF directly measures the temperature of the metastable ions, the measurement may differ from the actual ion temperature, depending on how long the metastable ions have to relax to their final temperature. The simulated FWHM of the IVDF as a function of the metastable lifetime is shown in Fig.~\ref{fig:f0qu}. It is clear that the LIF measurements of ion temperature only represent the typical ion temperature if the metastables have a sufficiently long lifetime. As their lifetime decreases, the measurement tends to become closer to the neutral temperature.

When an electrostatic wave is present in a plasma, the lifetime of metastable ions characterizes how long they typically experience the wave field. For the metastables that are produced from neutrals, they can only start to respond to the electric field once they are ionized. If the lifetime is too short compared to one wave period, these metastable ions will not live long enough to react to the wave, resulting in a reduction of the measured wave amplitude. Fig.~\ref{fig:f1f0} shows the simulated ratio of $f_1(v)$ to $f_0(v)$ at $v=v_\textup{t}$ as a function of the metastable lifetime when the electric field $\mathbf{E}$ is uniform in space. When the lifetime is longer than a wave period, $f_1/f_0$ approaches the correct value. However, this ratio drops as the lifetime becomes shorter. 

\subsection{Discussion}

\begin{figure}
\begin{center}
\includegraphics[width=3.2in]{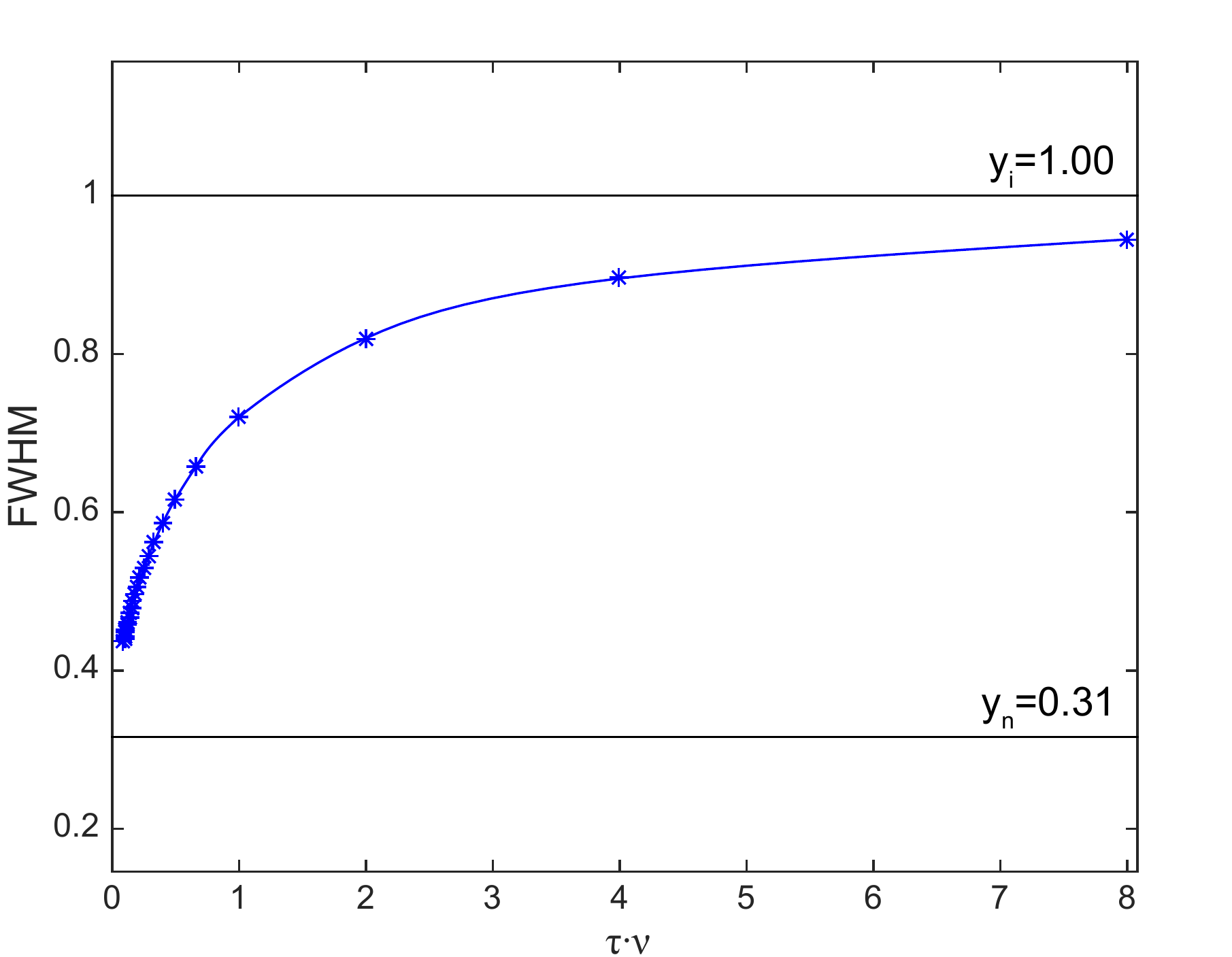}
\caption{Simulated FWHM of the IVDF as a function of the metastable lifetime $\tau$. $y_\textup{i}$ and $y_\textup{n}$ are the actual full widths of the ion and neutral particle velocity distribution respectively. All FWHM values in this plot are normalized by the actual full width of the ion velocity distribution. The LIF measured ion temperature is closer to the actual ion temperature when the metastable ions have a longer lifetime.}
\label{fig:f0qu}
\end{center}
\end{figure}
\begin{figure}
\begin{center}
\includegraphics[width=3.2in]{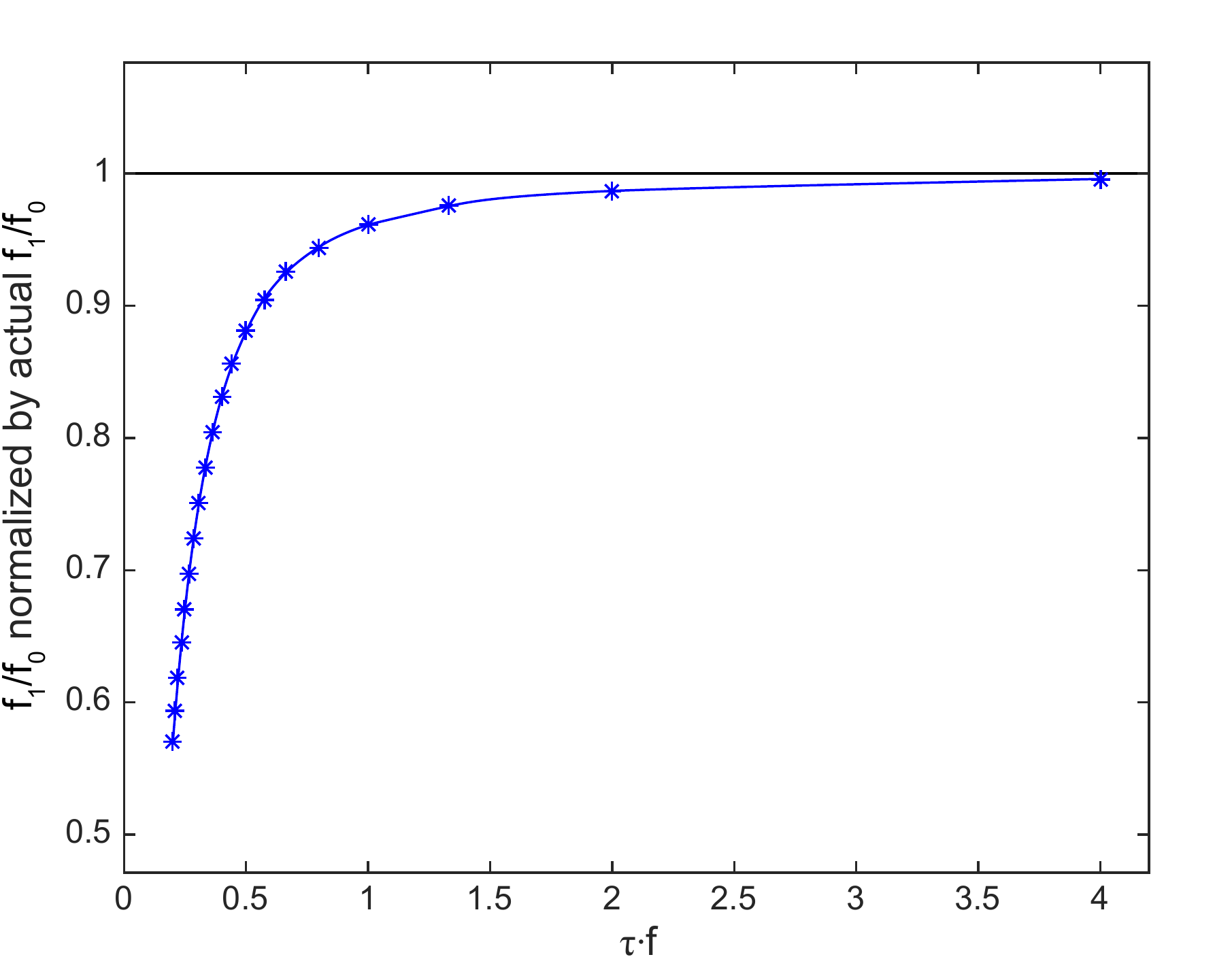}
\caption{Simulated ratio of $f_1(v)$ to $f_0(v)$ at $v=v_\textup{t}$ for various metastable lifetimes $\tau$ when electric field $\mathbf{E}$ is uniform. When the lifetime is longer than a wave period, $f_1/f_0$ is valid. However, this ratio drops rapidly when the lifetime is shorter than a wave period.}
\label{fig:f1f0}
\end{center}
\end{figure}

One important consequence of the metastable lifetime effects illustrated in Figs.~\ref{fig:f0qu} and \ref{fig:f1f0} is that under circumstances where the metastable ion population is coming directly from the ionization of neutrals (as opposed to the excitation of ground-state ions), the velocity distribution will only faithfully represent processes which act on the ion dynamics in a time shorter than the metastable lifetime. Under such circumstances, the perturbed distribution $f_1(v,t)$ cannot be correct if the wave frequency is lower than $1/\tau$ (inverse metastable lifetime). Similarly the ion-ion coulomb collision frequency must be higher than $1/\tau$ in order for the temperature to be correct. These results suggest that the contribution of these metastable ions to the LIF signal should be fundamentally interpreted as that of test particles. Although the electronic cross section for direct production of metastables from neutrals is significantly smaller than the excitation cross section that produces metastables from ground state ions (and generally requires higher energy electrons), the density of neutrals can be orders of magnitude higher than the ground state ion density. Evidence for this has been found in experiments \cite{Cooper}. 

As for the metastables produced from pre-existing ions, the results are less complicated. Given that these metastable ions have a history as ``typical'' ions and already represent the actual ion distribution, the measured FWHM of the IVDF is independent of metastable lifetime. Similarly, as these metastable ions have started to react to the wave field before they are produced, $f_1/f_0$ is also valid independent of metastable lifetime. However, the optical pumping broadening and coulomb collision effects shown in Secs. \ref{IVDF} and \ref{pert} are still expected to occur to this metastable population.

Overall, LIF can be a powerful tool in studying velocity distributions and plasma waves under the proper conditions. We have shown that some parameters, such as laser intensity, collision frequency, and metastable quenching, need to be considered when using LIF. With the Lagrangian model, the systematic effects due to these factors can be corrected.

\section{Summary}
\label{sec:summary}

In this paper, we report a general Lagrangian model by introducing a conditional probability function valid for many ion collision times. Numerical simulations show that optical pumping broadening affects measurement of the ion distribution function $f_0(v)$ and its perturbation $f_1(v,t)$ when laser intensity is sufficiently high. The results also show that the lifetime of metastable ions can affect the LIF diagnosis of ion temperature and electrostatic waves. They suggest that the IVDF measurements are only accurate when the ion-ion collision frequency is higher than the inverse metastable lifetime. As for wave detection, the wave frequency has to be significantly larger than the inverse metastable lifetime for a direct interpretation. Experiments are carried out to compare with the simulation in order to validate the model.

This Lagrangian approach includes a variety of physical processes that are important in LIF measurements, such as veclocy-space diffusion, time dependent optical pumping, and metastable quenching. The method permits further extension, provided that an appropriate conditional probability function can be constructed.


\begin{acknowledgments}

One of the authors (F. Chu) wishes to thank S. Mattingly, W. D. S. Ruhunusiri, J. Berumen and S. Kunhammed for helpful discussions. This work was supported by the U.S. Department of Energy under Grant No. DE-FG02-99ER54543. 

This research is part of a Ph.D. dissertation to be submitted by F. Chu to the Graduate College, University of Iowa, Iowa City, IA.

\end{acknowledgments}

\bibliography{refs}

\end{document}